\documentstyle[preprint,aps]{revtex}
\begin{document}
\draft
\title{Transitions in non-conserving models of Self-Organized Criticality}
\author{Stefano Lise$^a$ and  Henrik Jeldtoft Jensen$^b$}
\address{Department of Mathematics, Imperial College\\
 180 Queen's Gate, London SW7 2BZ\\
United Kingdom}
\maketitle
\begin{abstract}
We investigate a random--neighbours version
of the two dimensional non-conserving earthquake model of Olami, Feder and
Christensen
[Phys. Rev. Lett. {\bf 68}, 1244 (1992)]. We show both analytically and
numerically
that criticality can be expected even in the presence of dissipation. As the
critical
level of conservation, $\alpha_c$, is approached, the cut--off
of the avalanche size distribution scales as $\xi\sim(\alpha_c-\alpha)^{-3/2}$.
The transition from non-SOC
to SOC behaviour is controlled by
 the average branching ratio $\sigma$
 of an avalanche, which can thus be regarded as an order parameter of the
system.
The relevance of the results are discussed in connection to the
nearest-neighbours OFC model (in particular we analyse the relevance of
synchronization in the latter).
\end{abstract}
\vspace{1cm}
\narrowtext
In what is now a commonly referenced
paper\cite{bak}, Bak, Tang and Wiesenfeld
proposed that extensive
systems of many coupled elements naturally evolve toward a dynamical critical
state. They named this phenomenon Self--Organized Criticality and tested
the idea on a cellular automata, the so called sandpile model.
It was shown in that paper, and many others which followed\cite{sand}, that,
independently of the starting, configuration the sandpile model
evolves to a stationary state which does not possess any characteristic spatial
scale. The amplitude of the response of
the system to an external perturbation follows a power law distribution.

{}From a  theoretical point of view the main effort has been directed towards
understanding which are the  fundamental mechanisms that lead a system to be
self--organized critical.
It is well known, for instance, that for a particular class of sandpile models
a necessary condition is a conserving local dynamics\cite{manna,dhar}:
any amount of
dissipation introduces a characteristic length--scale and thus destroys
criticality.
Nonetheless this does not seem to be a universal requirement, since
in some other  classes of models,  conservation does not appear to be
needed.
In this respect the earthquake model introduced by Olami, Feder and
Christensen\cite{olami}
is of particular interest.  In this model it is possible to directly
 control
the level of conservation of the dynamics, through a parameter $\alpha$.
When $\alpha = 1/4$ the system is conserving and it probably belongs to the
same universality class as the sandpile model\cite{bak}.
But, contrary to the latter, there are converging evidences
that  the OFC model remains
critical even when dissipation is introduced
($\alpha < 1/4$)\cite{olami,kertez,socolar,grassberger,corral,middleton}.
By reducing the value of $\alpha$ one should be able in principle to observe
a transition from S.O.C. behaviour to non--S.O.C, although a general
agreement on  the critical value , $\alpha_c$, does not yet exist.
The spectrum of values ranges from the recently proposed
$\alpha_c=0$\cite{middleton}
to  $\alpha_c\simeq0.18$\cite{grassberger,corral}, passing through
the originally estimated
$\alpha_c\simeq0.05$\cite{olami,kertez} \cite{impossible}.
It has been suggested that the observed criticality in this  model is
a consequence of the ``imperfect'' synchronization in the system
\cite{middleton,kim-phd}.
The strong correlations so induced would allow an avalanche
(i.e. the response to an
external perturbation) to be infinitely big, although it keeps on losing
``energy''.

In this letter we investigate the random neighbours version of the OFC
model\cite{olami}. We find that a sharp transition from non-S.O.C.
to S.O.C. behaviour occurs at $\alpha_c\approx2/9$ in this model.
Even more interesting, this transition is
correctly described by the branching ratio, which can thus be
regarded as an order parameter of the system, in analogy to equilibrium
phase transitions.
The results obtained for the random neighbours model are interesting in their
own right, but their relevance is fully recognised when they are related to the
nearest neighbours version. In fact they provide not only a mean--field
description
of the OFC model, but also properly address the role played by synchronization.
For a system where the sole mechanism leading to SOC is synchronization,
one would expect the random neighbours version to be non critical,
since the latter cannot synchronize.

The OFC model is a a coupled-map lattice model which, despite its
simplicity, is thought to capture some of the essential features of earthquake
dynamics\cite{burridge}.
To each site of a $2$--dimensional square lattice is associated a real
continuous ``energy'' $E_{i}$.
The system is driven continuously and uniformly, which means that all the
$E_{i}$ values are simultaneously increased with time at the same rate.
Avalanche (or earthquake) dynamics is simulated by assuming that
a single site is unable to store more than a finite amount of energy $E_c$.
As soon as a site becomes unstable (i.e. $E_{i} \geq E_c$) an avalanche is
triggered: the global driving is stopped and the system evolves
according to the following local relaxation rule:
\begin{equation}
\label{eq.10}
\mbox{if $E_{i} \geq E_c \Rightarrow$} \left\{ \begin{array}{ll}
                 E_{i} \rightarrow & 0 \\
                  E_{nn} \rightarrow &
   E_{nn}+\alpha E_{i}
                 \end{array}
\right.
\end{equation}
until all of sites are below $E_c$. In eq.(\ref{eq.10}) ``$nn$''
stands for the  collection of nearest neighbours to site $i$.
The parameter  $\alpha \in [0,\frac{1}{4}] $ controls the conservation level
of the dynamics ($\alpha =\frac{1}{4}$ corresponds to the conservative case).
The random neighbours version we consider differs from the
OFC model only in the choice of neighbours: an unstable site distributes
an energy  $\alpha E_i$ to $4$  randomly chosen sites.
It is well known that boundary conditions play a crucial role in
SOC\cite{zapperi}.
In accordance with previous studies we use open boundary conditions
so that if one of the random neighbours is a boundary site the
 energy $\alpha E_i$ is simply lost.

The presence of a transition from localised to  non--localised behaviour
of the avalanches can be understood through the following argument.
For notational clarity in the following we will distinguish between a
stable and
an unstable site through the superscripts $-$ and $+$.
Let $P_+(E^+)$ be the probability that a random site will become active
as a consequence of receiving a contribution of magnitude
$\alpha E^+$.
$P_{+}(E^+)$ is equal to the probability of a
site having a
value between $E_c-\alpha E^+$ and $E_c$.
The average number of active sites produced by un unstable site
with energy $E^+$ is then $4 P_{+}(E^+)$.The branching ratio
$\sigma$ is defined as the average number of new active sites created
by an unstable site.
Averaging over the whole spectrum of  possible $P_+(E^+)$
(that is an average over $E^+$) we get for $\sigma$
\begin{equation}
\label{eq.30}
\sigma = 4 \int_{E_c}^\infty P_+(E^+)P(E^+)dE^+ \equiv 4 P_{+}
\end{equation}
where $P(E)$ is the distribution of the dynamical variable in the system.
Obviously the chance to have an infinite avalanche is directly related
to the condition $\sigma \geq 1$.   \\
An exact analytic calculation of $P_+$ is, unfortunately, too complicated,
since
it
involves a detailed knowledge of $P(E)$ (see fig.1).
In order to continue we have to assume a specific functional form for $P(E)$.
For simplicity we approximate the distribution of subcritical $E$-values by
a uniform distribution on the interval $[0,E_c]$. We obtain
$P_{+}(E^+) =  \alpha E^+/E_c$ and therefore
$P_{+}  = \alpha\langle E^+\rangle/E_c$
where $\langle E^+\rangle$ is the average value of a collapsing site.
Consequently it follows that
\begin{equation}
\label{eq.60}
\sigma = 4 \alpha \frac{\langle E^+\rangle}{E_c}
\end{equation}
It is immediately clear from eq. (\ref{eq.60}) that $\sigma$ will be greater
then $1$ even for some values of $\alpha < 1/4$ since
$\langle E^+\rangle > E_c$. The condition $\sigma>1$ can  also be read
as a conservation law.  It  states that the average
contribution $4\alpha\langle E^+\rangle$ to the avalanche from the collapse
of an unstable  site must be greater or equal
to $E_c$ otherwise the avalanche will die exponentially.

We now estimate $\langle E^+\rangle$.
Consider an active  site $i$ and a site $j$ which will
be active as a consequence of the action of $i$. We have
$E_j^+ = E_j^- +  \alpha E_i^+$, where the superscripts $+$ and $-$
in the site $j$ distinguish between two  successive time steps.
Taking the average on both side of this equation we get
$\langle E_j^+ \rangle = \langle E_j^- \rangle +
\alpha \langle E_i^+ \rangle$.
By assumption $E_c-\alpha E_i^+<E_j^-<E_c$. Thus, if we again approximate the
distribution of $E$-values of the subcritical sites by a uniform
distribution on $[0,E_c]$ we will have
$\langle E_j^-\rangle = E_c-\frac{1}{2}\alpha \langle E_i^+\rangle.$
Moreover, since $ E_j^+ $ is a new active site
$\langle E_j^+ \rangle =  \langle E_i^+ \rangle$.
Combining these expressions we obtain
\begin{equation}
\label{eq.1100}
\langle E^+ \rangle= \frac{E_c}{1-\frac{\alpha}{2}}
\end{equation}
The branching ratio is accordingly given by
\begin{equation}
\label{eq.120}
\sigma = \frac{4 \alpha}{1-\frac{\alpha}{2}}
\end{equation}
The condition for infinite avalanches  $\sigma \geq 1$ is then
\begin{equation}
\label{eq.130}
\alpha\geq \alpha_c = \frac{2}{9} \simeq 0.222...
\end{equation}
The above calculation, although very crude in the assumption
of a uniform distribution of subcritical $E$-values, predicts two
different phases for the random neighbours model: a ``low--$\alpha$'' phase,
where avalanches are essentially smaller than a characteristic size, and
a ``high--$\alpha$'' phase, where on the contrary it is possible to have
avalanches of any sizes.
Computer simulations indeed confirm this point.
Figure 2 shows the avalanche size distribution for increasing values
of $\alpha$.
The size is measured by counting the total number of toppling in one avalanche.
For $\alpha\leq0.22$ the distribution are very well described by the function
\begin{equation}
\label{eq.140}
P_{\alpha}(s) \propto s^{-3/2} \exp (\frac{s}{\xi})
\end{equation}
where the cut--off $\xi$, for sufficently large system, is independent of
$L$ ($L$ is linear dimension of the system). Fig. 3 shows the fit of the
simulated
$P_\alpha(s)$ distributions to the exponential form in Eq. \ref{eq.140}.
For $\alpha\geq 0.23$ the cut-off in $P_\alpha(s)$ scales with $L$.
This is the signature of criticality.
In fig.4 we show $\xi$ as a function of $\alpha$.
 The data fits well the expression
$\xi \sim (\alpha_c - \alpha)^{-1.5}$, with $\alpha_c = 0.2255$.
Note that $\xi\sim (1/4-\alpha)^{-2}$ was found in the study of the random
neighbour sand pile model\cite{kim-pre}, i.e. the coherence length is
only infinite in this model when conservation is established at $\alpha=1/4$.
Even more exciting is the behaviour of the measured branching ratio.
In fig.5 we report the measured branching ratio for different $\alpha$ as a
function of $1/L$. One sees that for $\alpha\leq 0.22$ the graphs extrapolate
to a value $\displaystyle{\lim_{L\rightarrow\infty}\sigma}$ well below 1. As
soon as
$\alpha\geq0.23$ $\displaystyle{\lim_{L\rightarrow\infty}\sigma\simeq 1}$ in
accordance
with
Eq. \ref{eq.130}.

In conclusion, by identifying a transition at a  finite conservation level
we have shown that a non-conserving system can indeed be critical.
The agreement between the analytical calculation and computer simulation
is remarkable and somewhat surprising. The assumption of a uniform
distribution of subcritical $E$-values is not a very accurate
approximation, see Fig.1. In this connection it is interesting to note
the observation made by Pietronero, Tartaglia, and Zhang\cite{pietronero}.
These authors found that the form of energy distribution depends
very much on the energy partition rule (i.e. on how the energy of
an unstable site is  distributed to its neighbours), whereas the
average energy in the system is quite universal.
This  is  a very interesting point, which would deserve further
investigations and  which might also explain why the uniform approximation
for  $P(E)$ gives such an accurate estimate of $\alpha_c$. It might be
possible  in fact that for a particular
energy partition rule the approximation becomes almost
correct (or even exact). In \cite{pietronero} was shown, for instance,
that a random partition makes  the peaks of Fig. 1 disappear.
We believe this property of universality might be related to the very nature
of S.O.C.:  measurable quantities should be independent of the details of
the model.
We have shown that synchronization is not the only mechanism present.
Rather, the criticality
appears to be related to the dynamics. This of course, does not prevent
synchronization to  be relevant for lower values of $\alpha$ as has
been suggested for the nearest neighbour version of the OFC
model \cite{middleton,kim-phd}.
Finally we have shown that the branching ratio plays the role of an order
parameter in the considered model.
We believe that this last point might also be useful for the analysis of
other models. The natural step is to use it for the nearest neighbour version
the OFC model. Preliminary results are
encouraging  even though simulations show that the scaling properties of
the branching ratio are more complicated than is the case in  the random
neighbour model.

{\it Acknowledgement} We are grateful to Peyman Ghaffari and Kent B.
Lauritsen for stimulating discussions.

\noindent a) email: s.lise@ic.ac.uk\\
b) email: h.jensen@ic.ac.uk\\

\begin{center}{\bf  Captions} \end{center}
\noindent{ \bf Figure 1.} \\
Energy distribution per site, $P(E)$, for $\alpha=0.23$ and  $L=400$
(continuous
line)
($E_c=1$).
The step function (dashed line) is the approximation  used in the calculations.
At $E=0$, $P(E)$ extends up to $\simeq 33$.\\
\noindent{ \bf Figure 2.} \\
Avalanche size distribution for $L=100$, $200$ and $400$ and for (a)
$\alpha=0.20$, (b)
$\alpha=0.21$, (c)~$\alpha=0.22$ and (d) $\alpha=0.23$.\\
\noindent{ \bf Figure 3.} \\
Avalanche size distributions for $\alpha < \alpha_c$. The ditributions do not
scale
with system size.
The continuous lines
are fits of the form $P_{\alpha}(s) \propto s^{-3/2} \exp (\frac{s}{\xi})$\\
\noindent{ \bf Figure 4.} \\
Cutoff in avalanche size distribution, $\xi$, as a function of
$\alpha_c-\alpha$
($\alpha_c=0.2255$).
The solid line is the interpolation $\xi \sim (0.2255 - \alpha)^{-1.5}$\\
\noindent{ \bf Figure 5.} \\
Measured branching ratio as a function of $1/L$ ($L=100$, $200$, $400$),
for different values of $\alpha$. From bottom to top, $\alpha=0.2,0.21,0.22,
0.225,0.23,0.24,0.25$. One can clearly see the abrupt change in behaviour for
$\alpha\simeq0.225$.

\begin{references}
\bibitem{impossible} To be complete, one should also add to this list
$\alpha_c=0.25$, which is the value suggested by  those who just don't
believe a  non--conserving system can be critical.
\end{references}
\end{document}